# Robust Set-Membership Diffusion Normalization Subband Adaptive Filtering Algorithms Over Distributed Networks

Yugang Han[1], Haiquan Zhao[2*], Tongge Fu[2], Dongyu Tan[2]

**Abstract:** With the development of wireless sensor networks, distributed networks have received widespread attention. According to the different ways of connecting the nodes in the distributed network can be divided into different structures, of which the diffusion type structure is the most commonly used one due to its simple, stable and reliable. In order to improve the robustness of the diffusion subband algorithm in distributed networks, the median absolute deviation (MAD) theorem is applied to the error boundary selection, and this paper proposes a diffusion subband algorithm with a robust boundary. Through simulations, it is verified that the proposed algorithm can effectively reduce the update step size in the face of outlier interference, so that the algorithm has a good convergence performance and also has good robustness to impulsive noise.

## 1. Introduction:

Distributed filtering achieves global estimation by sharing data among multiple nodes in a distributed network, which leads to more reliable estimation results than a single node [1]-[3]. Several traditional algorithms have been extended to distributed filtering, like distributed least-mean square algorithm (DLMS) [4], distributed normalization subband adaptive algorithm (DNSAF) [4], distributed affine projection algorithm (DAPA) [5]. Literature [6] applying improved multi-band structured subbands to distributed network and utilizing the input regressions of each subband to improve the speed of the algorithm. Literature [7] extends the range of algorithmic applications by applying distributed subband filtering to complex numbers. The above algorithms, although multiple nodes can communicate directly with each other to reduce the estimation error, are still not resistant to impulsive noise, and the performance of the algorithms will be greatly reduced in the face of non-Gaussian components generated by impulsive noise. Some scholars have applied M-estimate to the distributed filtering algorithm in order to improve its ability to cope with impulsive noise, and proposed the distributed normalized least mean M-estimate algorithm (DNLMM) [8], and diffusive maximum correlation correntropy algorithm based on maximum correlation entropy criterion (DMCC) [9][10]. However, this type of algorithm still has various limitations, such as being affected by signal correlation. Among the set-member filtering algorithms, some scholars are inspired by the MAD method in M-estimate [11]. An LMS algorithm with robust set-member error bounds is proposed (SM-NLMS-REB), the algorithm takes the error bounded data space consisting of all input-desired signals $(u,d)\in S$, and divide it into sub-spaces $S=S_a\cup S_b\cup S_c\cup S_d$, different sub-spaces contain different proportions of Gaussian and non-Gaussian signals, which are processed in different sub-spaces using different update rate, one for achieving faster convergence performance and the rest for suppressing impulsive noise interference. In this way, the algorithm can improve the convergence while obtaining the ability to resist impulsive noise, while some scholars have proposed a variety of robust set-member error boundary filtering algorithms with better performance like RSM-APA [12]、RSM-SAF [13]. Therefore, in this paper, we apply the set-element robust bounds to DNSAF and propose the set-member diffusion normalization subband adaptive filtering algorithm with robust boundary (RSM-DNSAF). Then we derive and design the update formula of the algorithm. Simulations show that the proposed algorithm has good robustness to impulsive noise, while improving the convergence speed of the algorithm.

[1] Yugang Han, Haiquan Zhao, Tongge Fu and Dongyu Tan are with the Key Laboratory of Magnetic Suspension Technology and Maglev Vehicle, Ministry of Education, and the School of Electrical Engineering, Southwest Jiaotong University, Chengdu610031, China. (email:yuganghan_swjtu@126.com; hqzhao_swjtu@126.com; tonggefu@126.com; dongyutan_swjtu@126.com)

*Corresponding author: Haiquan Zhao.

## 2. RSM-DNSAF derivation

Consider a diffusion network with N nodes, where the signal is segmented among each node using multiplicative subband structure, defining the linear model for a n-node as follows:

$$d_n(l)=\mathbf{u}_n^T(l)\mathbf{w}_o+\eta_n(l) \tag{1}$$

where $\mathbf{u}_n^T(l)$ is the transpose of the input signal vector of node n at moment $l$, $\mathbf{w}_o$ is a vector of unknown system weights, its order is L-dimensional, $\eta_n(l)$ is background noise, which is Gaussian white noise with variance $\sigma_{v,n}^2$, $d_n(l)$ is the desired output signal corresponding to the unknown system.

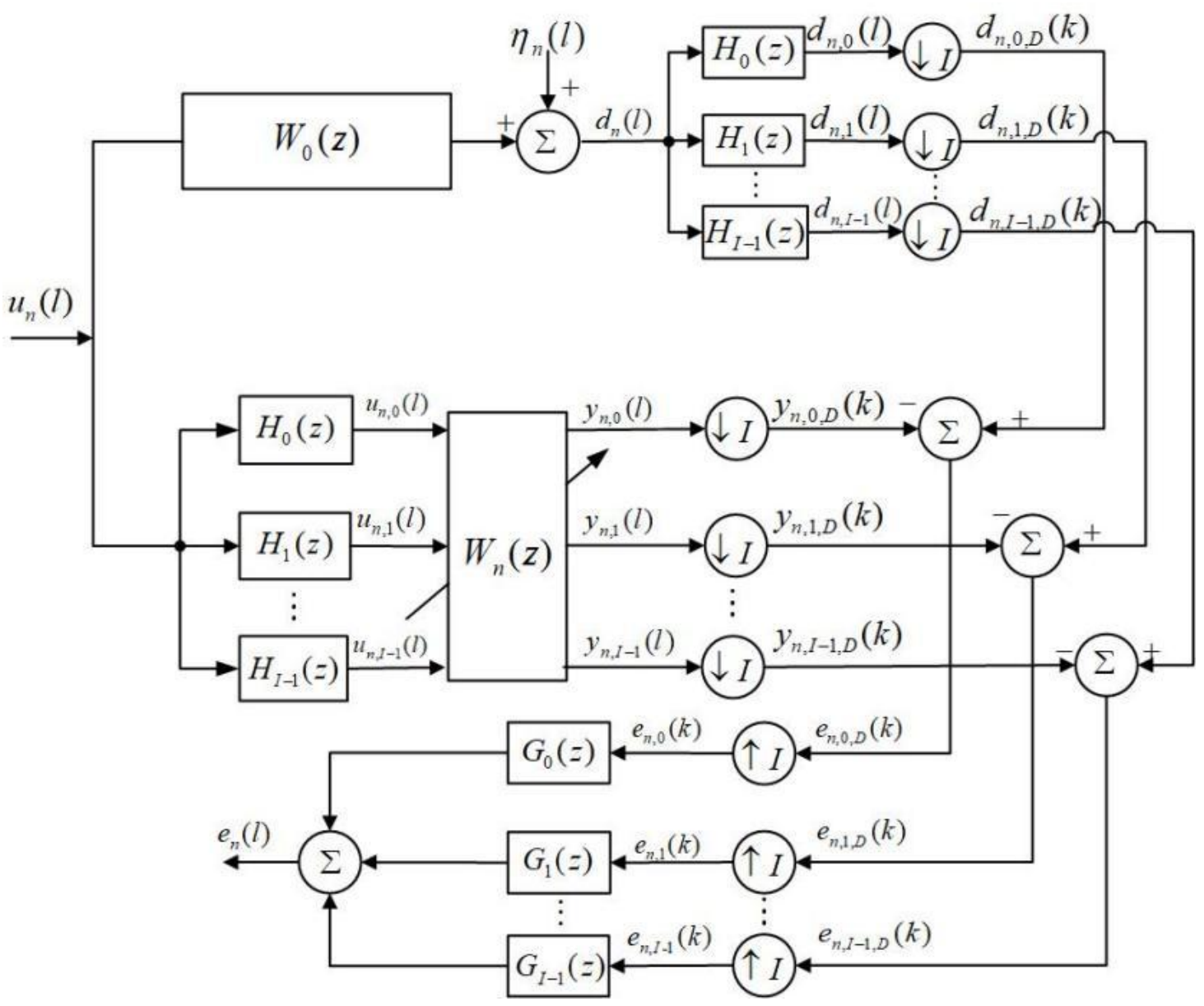

Fig 1 Subband filters for each node of the distributed network

Fig 1 denotes the full band structure in a single node. The input signal is divided into subband signals by a decomposition filter ban. Then, we can get $d_{n,i,D}(k)$, $y_{n,i,D}(k)$ and $e_{n,i,D}(k)$, $i$ denotes the i-th subband, k denotes the extracted time series.

Distributed subband filtering differs from the traditional subband filtering in that the results estimated by a single node can only be used as an intermediate weight vector. Defining the local estimation of n nodes as $\mathbf{h}_n(k) \in R^{L\times 1}$, then we can get the following optimization problem for set-member diffusion subband filtering on a single node:

$$\begin{aligned} &\min_{\mathbf{h}_n(k)} \ \| \mathbf{h}_n(k) - \mathbf{w}_n(k-1) \|_2^2 \\ &\text{s.}t \ \ \mathbf{d}_{n,D}(k) - \mathbf{U}_n^T(k)\mathbf{h}_n(k) = \mathbf{g}_n(k) \end{aligned} \tag{2}$$

where $\mathbf{d}_{n,D}(k) \in R^{I\times 1}$ denotes the desired output signal vector of node $n$, $I$ is the number of subbands. $\mathbf{U}_n(k) = [\mathbf{u}_{n,0}(k), \mathbf{u}_{n,1}(k), ..., \mathbf{u}_{n,I-1}(k)] \in R^{L\times I}$ denotes the input signal matrix at node $n$ at moment $k$, $\mathbf{h}_n(k)$ denotes the intermediate weight vector of node $n$ at moment $k$, $\mathbf{g}_n(k) = [g_{n,0}(k), g_{n,1}(k), ..., g_{n,I-1}(k)]^T \in R^{I\times 1}$ denotes the error boundary vector. Solving the above constrained optimisation problem using the Lagrange multiplier method yields the following cost function.

$$f(k) = \| \mathbf{h}_n(k) - \mathbf{w}_n(k-1) \|_2^2 + \left[ \mathbf{d}_{n,D}(k) - \mathbf{U}_n^T(k)\mathbf{h}_n(k) - \mathbf{g}_n(k) \right]^T \boldsymbol{\theta}_n \tag{3}$$

where $\boldsymbol{\theta}_n(k) = \left[ \theta_0(k), \theta_1(k), ..., \theta_{I-1}(k) \right]^T$ is a Lagrange multiplier vector.Using the gradient descent method and the weight update formula can be obtained as follows:

$$\mathbf{h}_n(k)=\begin{cases}\mathbf{w}_n(k-1)+\mu_n\sum_{i=0}^{I-1}\frac{\mathbf{u}_{n,i}(k)}{\|\mathbf{u}_{n,i}(k)\|_2^2}(e_{n,i,D}(k)-g_{n,i,D}(k)) & if\ |e_{n,i,D}(k)|>\gamma\\ \mathbf{w}_n(k-1) & otherwise\end{cases} \tag{4}$$

where $e_{n,i,D}(k)=d_{n,i,D}(k)-\mathbf{u}_{n,i}^T(k)\mathbf{w}_n(k-1)$ are a priori errors, defining the subband error boundary as $g_{n,i}(k)=\gamma\, sign(e_{n,i,D}(k))$ , $sign(\cdot)$ denote a symbolic function, $\gamma=\sqrt{t\sigma_{v,n}^2}$ as the specified boundary of the subband error signal, which can be used as a judgement of whether the weight vector is updated or not. Define the set of constraints of $\mathbf{w}_n(k)$ as follows, according to set member filtering theory.

$$\begin{gathered}\mathrm{w}_n(k)\subset(\psi_{n,k,0},\psi_{n,k,1}\cdots\cdots\psi_{n,k,I-1})\\ \psi_{n,k,i}=\left\{\mathbf{w}_n\in\mathrm{R}^L:\ \left|d_{n,i,D}(k)-\mathbf{u}_{n,i}^T(k)\mathbf{w}_n\right|\le\gamma, i\in[0,I-1]\right\}\end{gathered} \tag{5}$$

Meanwhile, using the combination coefficients between n nodes and other nodes, the local estimations are combined to form a global estimation at moment $k$. The update formula for the set-member diffusion subband algorithm in adapt-then-combine (ATC) mode can be obtained as:

$$\begin{cases}\mathbf{h}_n(k)=\mathbf{w}_n(k-1)+\mu_n\sum_{i=0}^{I-1}\alpha_{n,i}(k)\frac{\mathbf{u}_{n,i}(k)}{\|\mathbf{u}_{n,i}(k)\|_2^2}e_{n,i,D}(k)\\ \mathbf{w}_n(k)=\sum_{m\in N_n}c_{m,n}\mathbf{h}_n(k)\end{cases} \tag{6}$$

$$\alpha_{n,i}(k)=\begin{cases}1-\frac{\gamma}{\left|e_{n,i,D}(k)\right|} & if\ \left|e_{n,i,D}(k)\right|>\gamma\\ 0 & otherwise\end{cases} \tag{7}$$

In the ATC mode, the $c_{m,n}$ are taken as Metropolis rules [14] as follows:：

$$c_{m,n}=\begin{cases}\frac{1}{\max(n_m,n_n)}, & if\ m\in N_n, m\ne n\\ 1-\sum_{m\ne k}c_{m,n}, & if\quad m=n\\ 0, & otherwise\end{cases} \tag{8}$$

$m\in N_n$ denotes that m and n are two nodes connected to each other.

In order to realize the robustness of the algorithm, the MAD criterion [12]-[13] is introduced into the setup of the set member boundaries [15]-[19], the $\gamma$ is：

$$\gamma=\frac{\left|e_{n,i,D}(k)\right|^2}{v\xi(k)+\left|e_{n,i,D}(k)\right|} \tag{9}$$

where $0 < v < 1$ , $\xi(k) = \tau\sqrt{\hat{\sigma}_e^2(k)}$ , $\hat{\sigma}_e^2(k)$ is given by:

$$\hat{\sigma}_e^2(k) = \lambda_\sigma \hat{\sigma}_e^2(k-1) + (1-\lambda_\sigma) med[\mathbf{A}_e(k)] \tag{10}$$

$\lambda_\sigma$ is a forgetting factor, $\hat{\sigma}_e^2(k)$ is a impulsive-free noise error signal, $med[\cdot]$ is to take the median, $\mathbf{A}_e(k) = [e_{n,i,D}^2(k)..e_{n,i,D}^2(k-N_w+1)]$ , $N_w$ is the length of the error signal sample. From equation (9), it can be seen that $|e_{n,i,D}(k)|$ is greater than $\gamma$ at any moment. However, this causes the algorithm to be updated all the time, which is contrary to the original intent of set-member filtering. So we assume that when the algorithm is in steady state, the actual error signal is equal to the error signal screened by the MAD theorem, for example $|e_{n,i,D}(k)| \approx \xi(k)$ , Then equation (9) becomes $\gamma \approx \frac{|e_{n,i,D}(k)|}{v+1}$ , comparing this with the regular set-member boundary, when $|e_{n,i,D}(k)| < \sqrt{t\sigma_{v,n}^2}$ , the error is too small to update. So equation (9) can be updated to:

$$\gamma = \begin{cases} \frac{\sqrt{t\sigma_{v,n}^2}}{v+1} & \frac{|e_{n,i,D}(k)|^2}{v\xi(k) + |e_{n,i,D}(k)|} < \frac{\sqrt{t\sigma_{v,n}^2}}{v+1} \\ \frac{|e_{n,i,D}(k)|^2}{v\xi(k) + |e_{n,i,D}(k)|} & otherwise \end{cases} \tag{11}$$

Substituting equation (11) into equation (6) and (7), the weight update formula for the RSM-DNSAF algorithm is obtained as:

$$\begin{cases} \mathbf{h}_n(k) = \mathbf{w}_n(k-1) + \mu_n \sum_{i=0}^{I-1} \alpha_{n,i}(k) \frac{\mathbf{u}_{n,i}(k)}{\|\mathbf{u}_{n,i}(k)\|_2^2} e_{n,i,D}(k) \\ \mathbf{w}_n(k) = \sum_{m \in N_n} c_{m,n} \mathbf{h}_n(k) \end{cases}$$
$$\alpha_{n,i}(k) = \begin{cases} \frac{v\xi(k)}{v\xi(k) + |e_{n,i,D}(k)|} & if \ |e_{n,i,D}(k)| > \sqrt{t\sigma_{v,n}^2} / (v+1) \\ 0 & otherwise \end{cases} \tag{12}$$

The variation of the robust setter bound with the error size is shown schematically in Figure 2. Where $c = \sqrt{t\sigma_{v,n}^2}$ . It can be seen that the proposed robust setter boundary will always be greater than or equal to the traditional setter boundary, and such a characteristic can make the setter algorithm have a lower computational amount, and also make it effective in improving the convergence speed of the algorithm and realizing robustness in the face of the anomalous noise due to the error signal and the

MAD criterion [20]-[23].

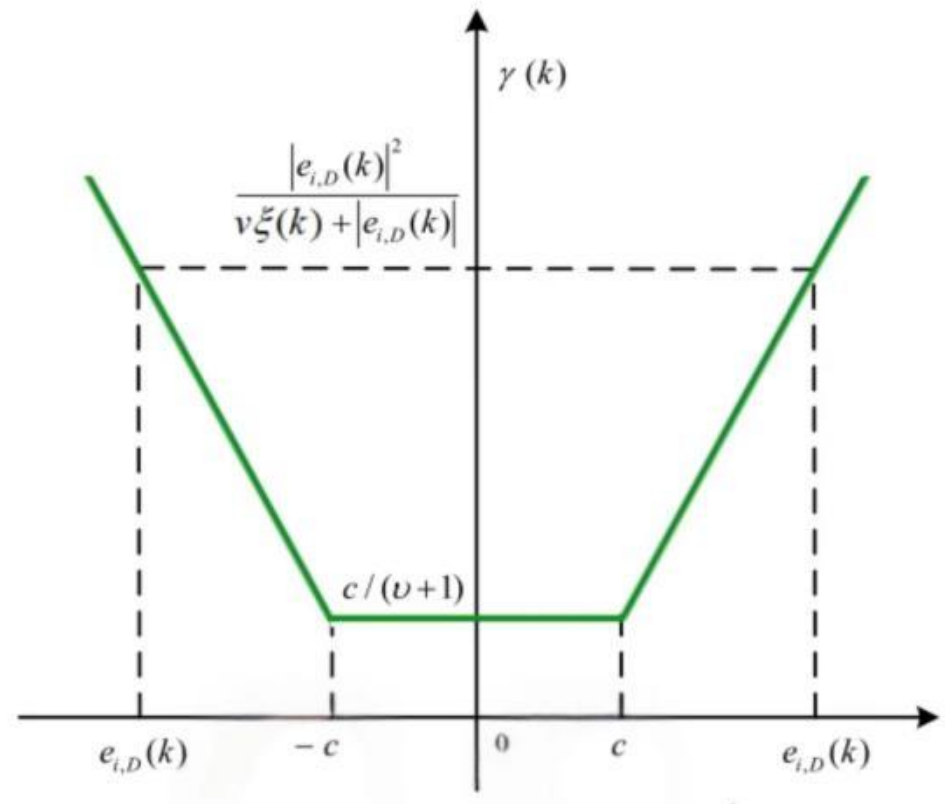

Fig 2 The curves for the robust set-member boundary $\gamma(k)$

## 3. Stability analysis

As with the linear model defined above, let the desired output signal at node $n$ be $d_n(l) = \mathbf{u}_n^T(l)\mathbf{w}_o + \eta_n(l)$ , $\mathbf{w}_o$ is a vector of unknown system weights. To analyze the stability of RSM-DNSAF, we define the following vector of weight deviations:

$$\overline{\mathbf{w}}(k) = \mathbf{w}_o - \mathbf{w}(k) \tag{13}$$

Combined with the RSM-DNSAF update formula, equation (14) is obtained by subtracting $\mathbf{w}_o$ from both sides of the equation (12) simultaneously:

$$\overline{\mathbf{w}}(k+1) = \sum_{m \in N_n} c_{m,n} \overline{\mathbf{w}}(k) - \mu_n \sum_{m \in N_n} c_{m,n} \sum_{i=0}^{I-1} \alpha_{n,i}(k) \frac{\mathbf{u}_{n,i}(k) e_{n,i,D}(k)}{\left\| \mathbf{u}_{n,i}(k) \right\|_2^2} \tag{14}$$

where

$$\alpha_{n,i}(k) = \begin{cases} \dfrac{v\xi(k)}{v\xi(k) + \left| e_{n,i,D}(k) \right|} & if \ \left| e_{n,i,D}(k) \right| > \sqrt{t\sigma_{v,n}^2} / (v+1) \\ 0 & otherwise \end{cases} \tag{15}$$

Multiply both the left and right sides of Eq.(13) together by $\mathbf{U}^T(k)$ . Then we can get:

$$\mathbf{e}_{n,i,D}(k) = \mathbf{U}^T(k)\overline{\mathbf{w}}(k) + \eta_n(k) \tag{16}$$

Also, define a transfer matrix:

$$\boldsymbol{C} = \begin{bmatrix} c_{1,1} & c_{1,2} & \cdots & c_{1,N} \\ c_{2,1} & c_{2,2} & \cdots & c_{2,N} \\ \vdots & \vdots & \vdots & \vdots \\ c_{N,1} & c_{N,2} & \cdots & c_{N,N} \end{bmatrix} \tag{17}$$

$$\boldsymbol{G} = \boldsymbol{C} \otimes \boldsymbol{I}_L \tag{18}$$

Combining equation (17) with equation (18), then step size set to 1, equation (14) can be simplified as:

$$\begin{aligned}\overline{\mathbf{w}}(k+1) &= \mathbf{G}\overline{\mathbf{w}}(k) - \mathbf{G}\boldsymbol{\alpha}(k)\frac{\mathbf{U}(k)}{\mathbf{U}(k)^T\mathbf{U}(k)}[\mathbf{U}^{\mathbf{T}}(k)\overline{\mathbf{w}}(k) + \eta_n(k)] \\ &= \mathbf{G}[I - \boldsymbol{\alpha}(k)\frac{\mathbf{U}(k)\mathbf{U}^{\mathbf{T}}(k)}{\mathbf{U}(k)^T\mathbf{U}(k)}]\overline{\mathbf{w}}(k) - \mathbf{G}\boldsymbol{\alpha}(k)\frac{\mathbf{U}(k)}{\mathbf{U}(k)^T\mathbf{U}(k)}\eta_n(k)\end{aligned} \tag{19}$$

By solving both sides of equation (19) simultaneously for the expected value, Equation (20) can be obtained:

$$\begin{aligned}E\left(\overline{\mathbf{w}}(k+1)\right) &= E\left(\mathbf{G}\overline{\mathbf{w}}(k) - \mathbf{G}\boldsymbol{\alpha}(k)\frac{\mathbf{U}(k)}{\mathbf{U}(k)^T\mathbf{U}(k)}[\mathbf{U}^{\mathrm{T}}(k)\overline{\mathbf{w}}(k) + \eta_n(k)]\right) \\ &= E\left(\mathbf{G}[I - \boldsymbol{\alpha}(k)\frac{\mathbf{U}(k)\mathbf{U}^{\mathrm{T}}(k)}{\mathbf{U}^T(k)\mathbf{U}(k)}]\overline{\mathbf{w}}(k)\right) - E\left(\mathbf{G}\boldsymbol{\alpha}(k)\frac{\mathbf{U}(k)}{\mathbf{U}(k)^T\mathbf{U}(k)}\eta_n(k)\right)\end{aligned} \tag{20}$$

Since the output noise $\eta_n(k)$ is independent of the input signal and is expected to be zero, the above equation can be simplified as

$$E\left(\overline{\mathbf{w}}(k+1)\right) = E\left(\mathbf{G}[I - \boldsymbol{\alpha}(k)\frac{\mathbf{U}(k)\mathbf{U}^{\mathrm{T}}(k)}{\mathbf{U}^T(k)\mathbf{U}(k)}]\overline{\mathbf{w}}(k)\right) \tag{21}$$

If the proposed algorithm is to be stable, the following equation is to be satisfied:

$$\| \overline{\mathbf{w}}(k+1) \|_2 < \| \overline{\mathbf{w}}(k) \|_2 \tag{22}$$

$\|\cdot\|_2$ denotes the 2-paradigm number, according to the maximum-paradigm inequality : $\| AB \|_2 < \| A \|_2 \| B \|_2$ . We can transform equation (21) into:

$$\begin{aligned}\| E\left(\overline{\mathbf{w}}(k+1)\right) \|_2 &= \left\| E\left(\mathbf{G}[I - \boldsymbol{\alpha}(k)\frac{\mathbf{U}(k)\mathbf{U}^{\mathrm{T}}(k)}{\mathbf{U}^T(k)\mathbf{U}(k)}]\overline{\mathbf{w}}(k)\right) \right\|_2 \\ &= \left\| \mathbf{G}[I - \boldsymbol{\alpha}(k)\frac{\mathbf{U}(k)\mathbf{U}^{\mathrm{T}}(k)}{\mathbf{U}^T(k)\mathbf{U}(k)}]E\left(\overline{\mathbf{w}}(k)\right) \right\|_2\end{aligned} \tag{23}$$

Just need to fulfill the $\sup\limits_{k\geq 0}\{\|\mathbf{G}\|_2 \cdot \left\| I - \boldsymbol{\alpha}(k)\frac{\mathbf{U}(k)\mathbf{U}^{\mathrm{T}}(k)}{\mathbf{U}^T(k)\mathbf{U}(k)} \right\|_2 \} < 1$ , which can stabilize the algorithm.Because of the convex combination used for the combination coefficients, the $\|\mathbf{G}\|_2 \leq 1$ . Meanwhile, as:

$$\alpha_{n,i}(k)=\begin{cases}\dfrac{v\xi(k)}{v\xi(k)+\left|e_{n,i,D}(k)\right|} & if \ \left|e_{n,i,D}(k)\right|>\sqrt{t\sigma_{v,n}^{2}}/(v+1) \\ 0 & otherwise\end{cases} \tag{24}$$

And because $\left|e_{n,i,D}(k)\right|\geq v\xi(k)$, so $0<\alpha_{n,i}(k)<1$, $\left\|I-\boldsymbol{\alpha}(k)\dfrac{\mathbf{U}(k)\mathbf{U}^{\mathrm{T}}(k)}{\mathbf{U}^{T}(k)\mathbf{U}(k)}\right\|_{2}<1$. Therefore, the proposed RSM-DNSAF algorithm is capable of stable convergence.

## 4. Simulation

In this subsection, simulation are conducted on the proposed algorithm to verify the performance. Using $\mathrm{NMSD}=\dfrac{1}{N}\sum_{n=1}^{N}MSD_{k}(l)$ as algorithm performance metrics and the regularization parameter is taken as 0.01.The input signals for each node are generated using through the auto-regressive systems AR (1) i.e. $x_{k}=0.95x_{k-1}$ and AR (4) i.e. $x_{k}=0.95x_{k-1}+0.19x_{k-2}+0.09x_{k-3}-0.5x_{k-4}$. The signal samples were all 40,000 in length and the background noise described above was added to each node. Using a 32-order unknown system with the subband filter order set to 4 and the expected weights vector $\mathbf{w}_{o}$ is generated by a random uniform distribution from 0 to 1. The number of analytic filter orders is the same as the number of subbands. Error signal sample length $N_{w}=9$, $\tau=2.576$, $\lambda_{\sigma}=0.99$, $v=0.5$. The combinational coefficients for the distributed network are generated using the Metropolis rule [14]. To ensure fairness, the experiments choose appropriate step sizes so that each algorithm has a same initial convergence rate, and all experimental results are averaged over 100 independent runs.

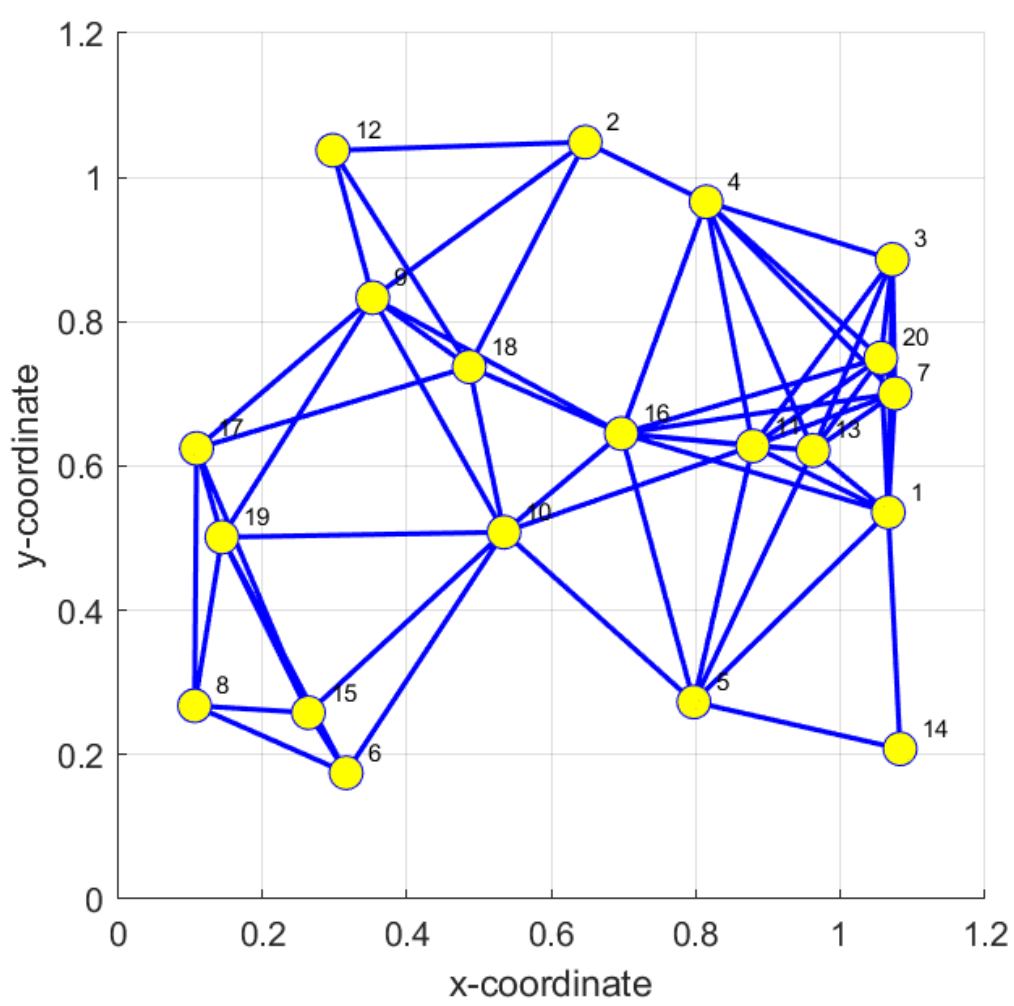

Fig 3 Distributed network topology with 20 nodes

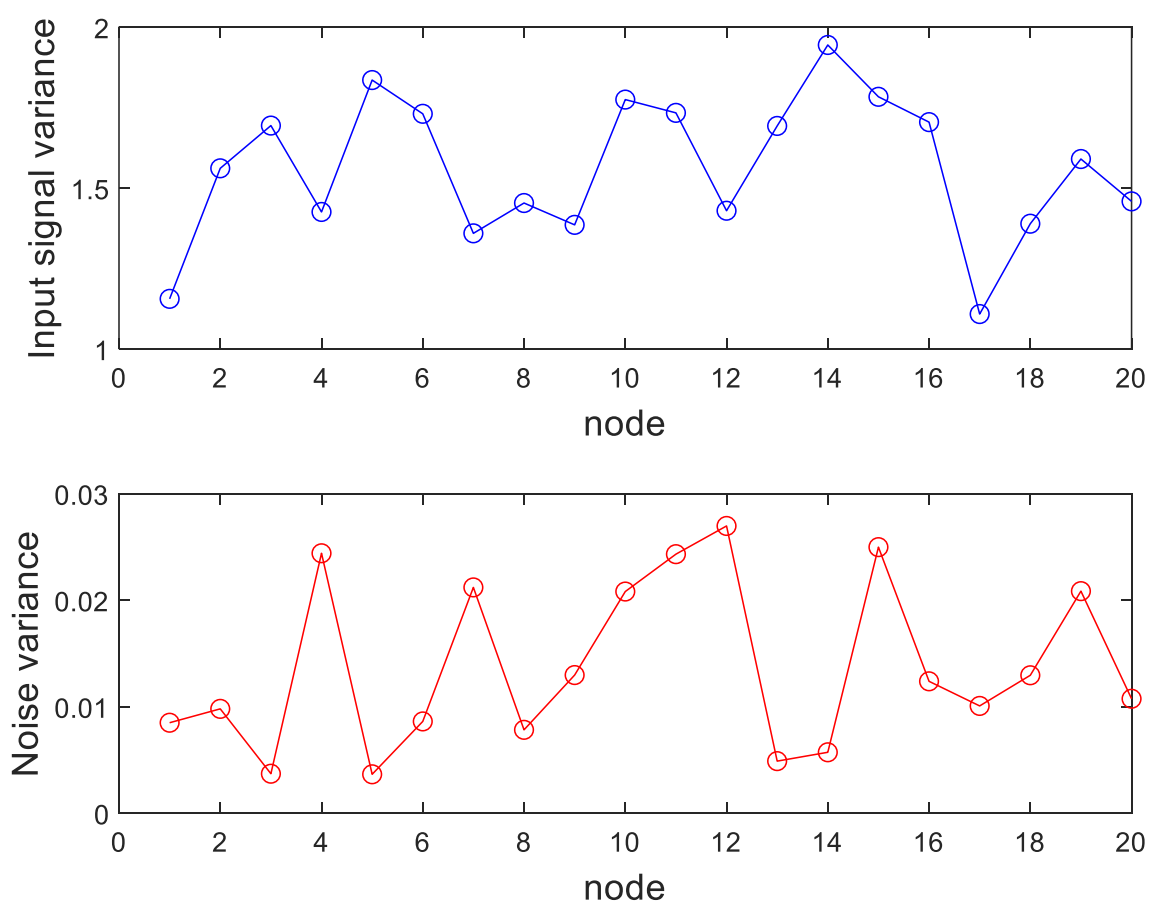

Fig 4 The value of $\sigma_{u,n}^2$ and $\sigma_{v,n}^2$ ,at 20 nodes

*A. 20 node inputs*

Considering a diffusion distributed network of 20 nodes first. The network topology is shown in Figure 3. Figure 4 shows the input signal and background noise variance data for each node. The step size of the SSAF algorithm is set to 0.8, and the convergence step of the RSM-NLMS, RSM-NSAF, SM-DNSAF and RSM-DNSAF algorithms are all set to 1. Impulsive noise occurs with probability $\Pr = 0.01$ and variance $\sigma_{i,n}^2 = 1000\sigma_{u,n}^2$ ,where $\sigma_{u,n}^2$ denotes the variance of the input signal. From Figures 5 and 6, it can be observed that due to the introduction of the robust set-member boundary, the set-member algorithm is robust and still achieves good filtering effect in the face of the impulsive noise. While the SM-DNSAF algorithm, which has not been enhanced with robustness, can no longer converge under the influence of the impulse noise and exhibits a divergence state. Moreover, compared with competing algorithms, the convergence of the proposed RSM-DNSAF algorithm is better than others, which enhances the applicability of the SM-DNSAF algorithm in distributed networks.

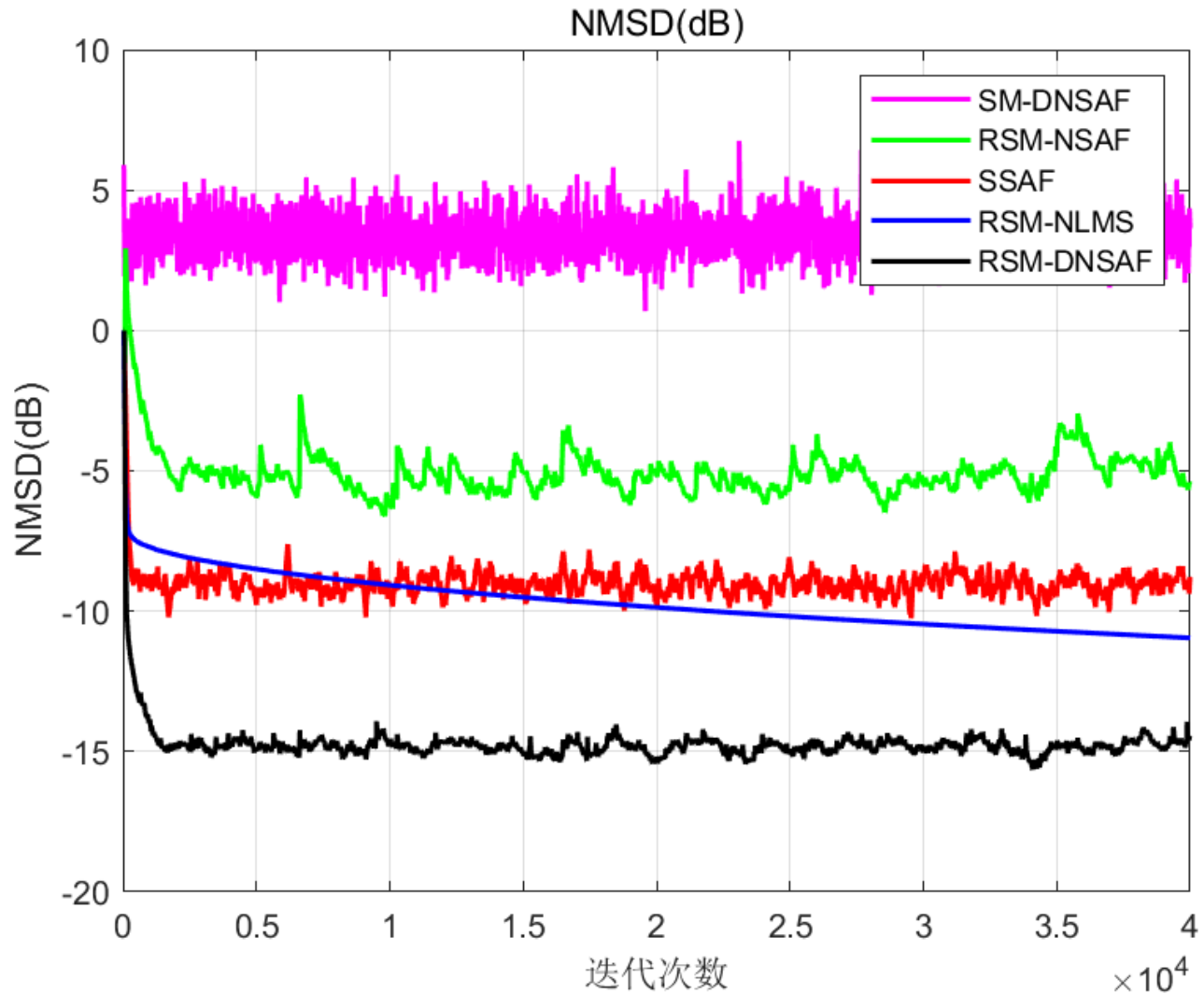

Fig.5 NMSD curves when the AR(1) input signal contains impulsive noise

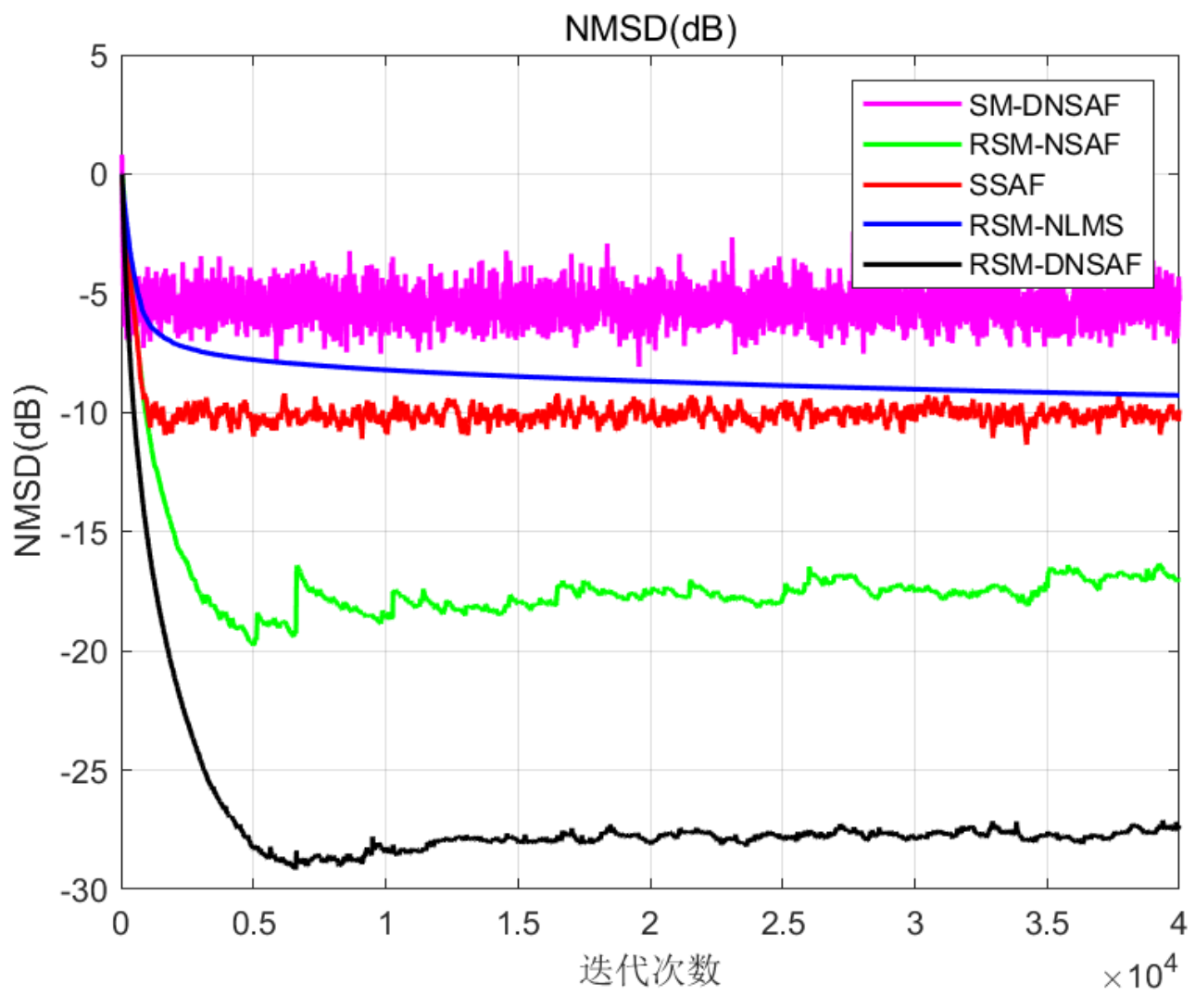

Fig.6 NMSD curves when the AR(4) input signal contains impulsive noise

### *B. 30 node inputs*

In the distributed network of 30 nodes, all the parameter settings are kept the same as in the 20-node distributed network. Fig.7 shows the distributed network topology with 30 information nodes, the input signal variance and background noise variance at each node are shown in Fig.8. From the convergence curves in Fig.9 and Fig.10, it is easy to see that the proposed RSM-DNSAF algorithm has good convergence and impulsive noise resistance in both AR(1) and AR(4) signals, which proves once again that the robust set-member boundaries approach improves the robustness of the SM-DNSAF algorithm in distributed networks.

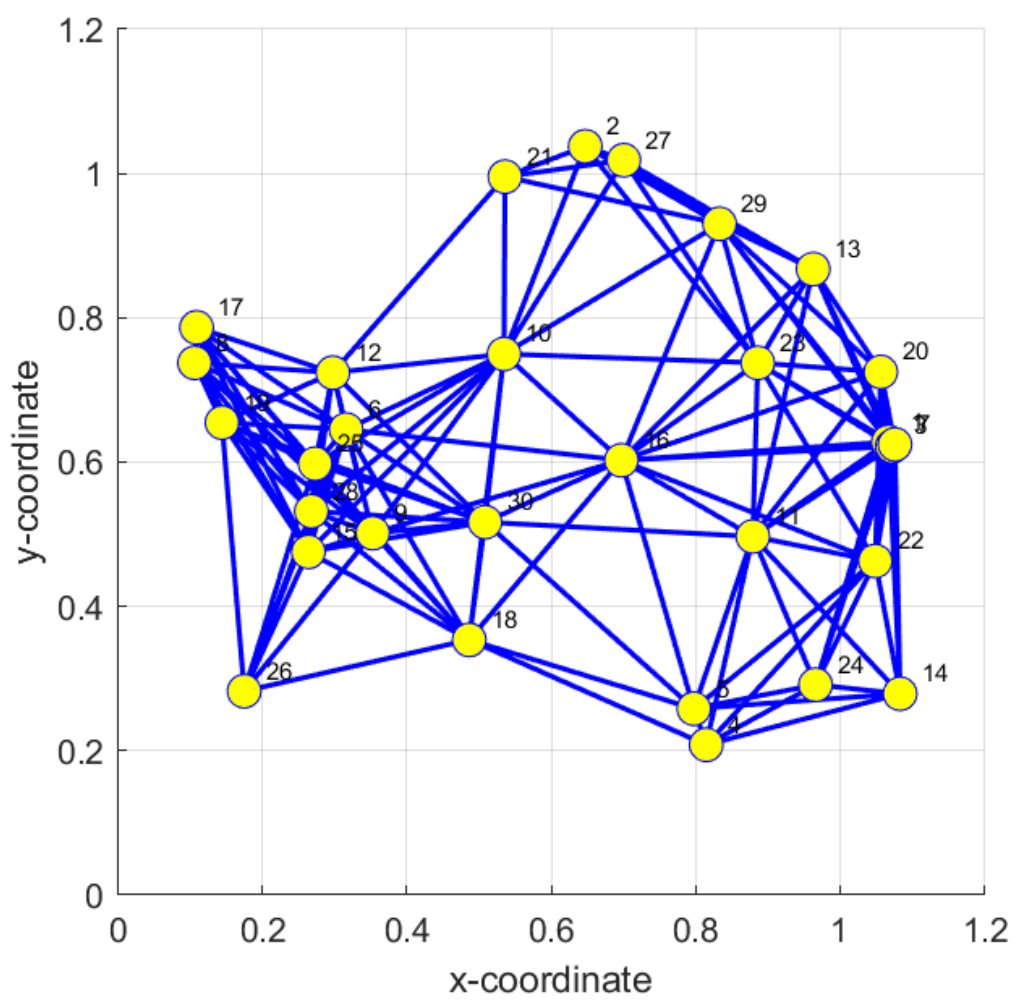

Fig 7 Distributed network topology with 30 nodes

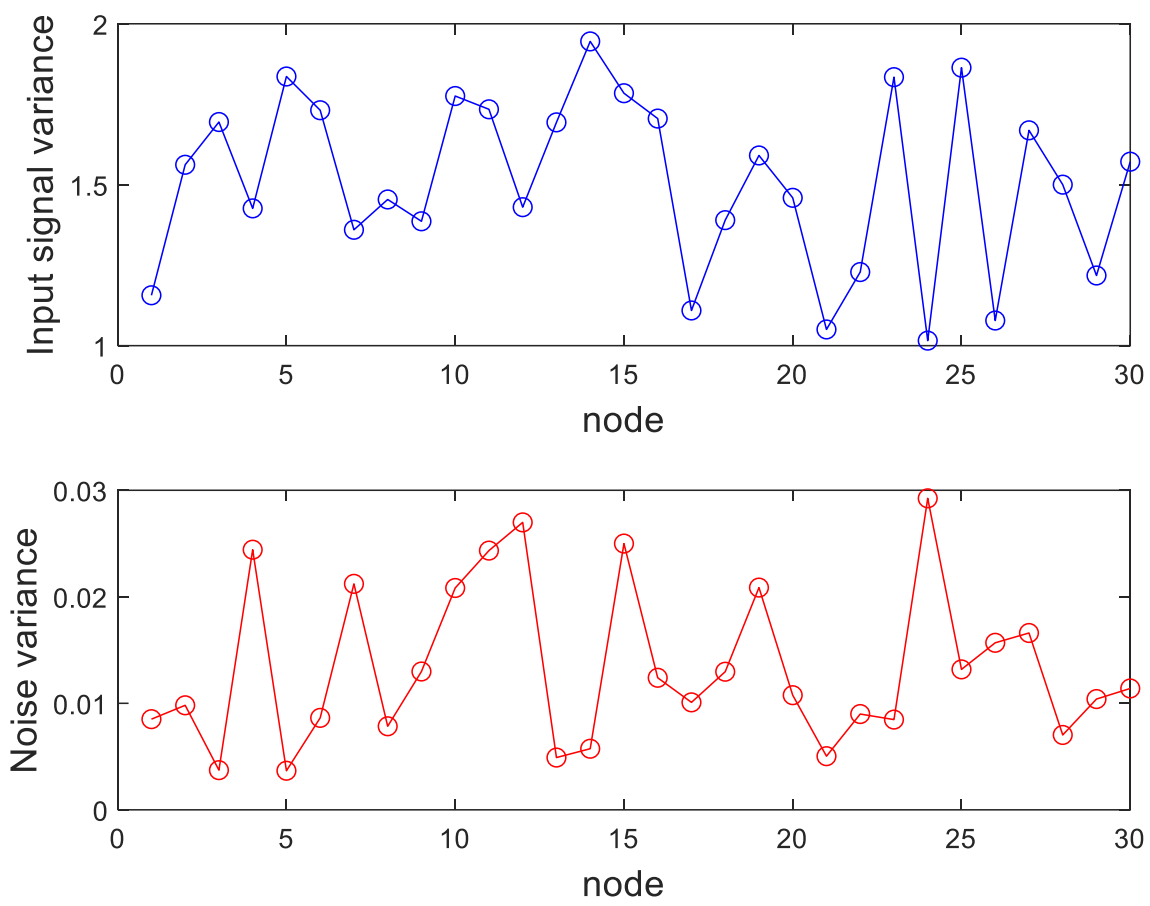

Fig 8 The value of $\sigma_{u,n}^2$ and $\sigma_{v,n}^2$ ,at 30 nodes

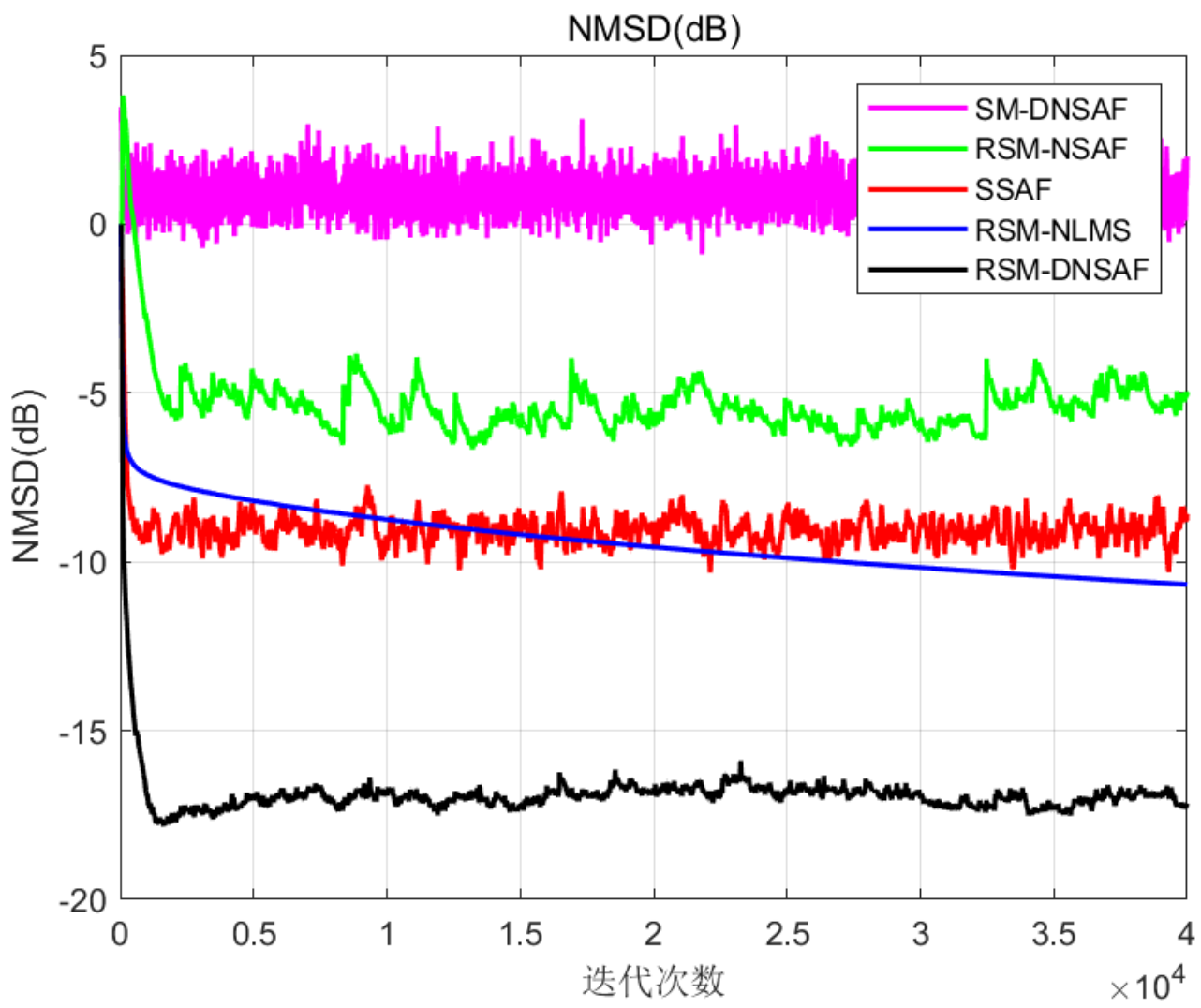

Fig 9 NMSD curves when the AR(1) input signal contains impulsive noise

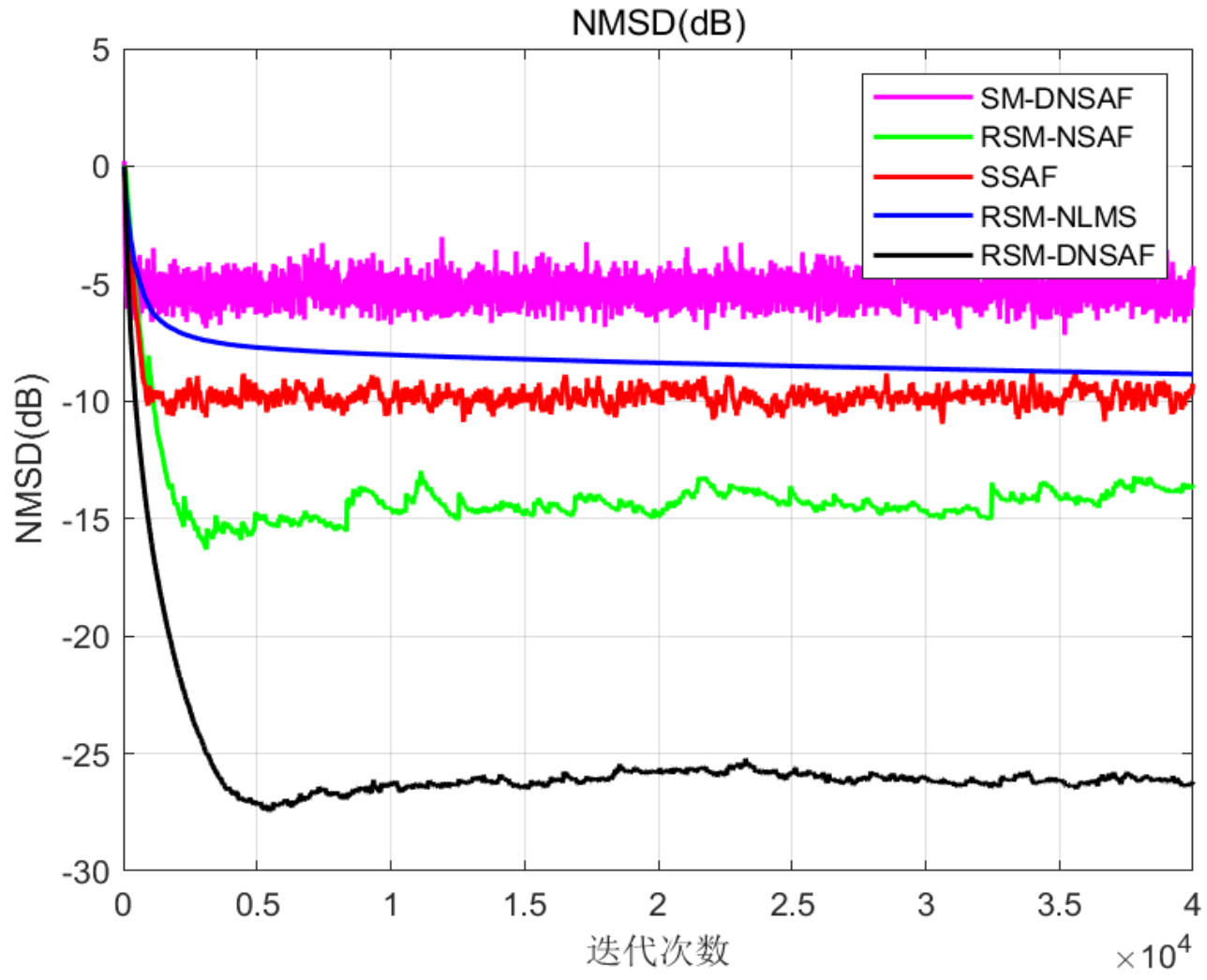

Fig 10 NMSD curves when the AR(4) input signal contains impulsive noise

## 5. Conclusion

In this paper, in order to improve the robustness of the diffusion subband algorithm, a new set-member boundary method is used and the set-member diffusion normalization subband adaptive filtering algorithm with robust boundary (RSM-DNSAF) is proposed. When impulsive noise interference occurs, the anomalous signal can slow down the updating speed of the weight vectors, thus achieving good suppression of impulse noise. Finally, the algorithm is simulated and verified by Matlab under different input signals and different distributed networks, and the experimental results show the superiority of the proposed algorithm.

### Data availability

The data that support the findings of this study are available from the corresponding author on request.

### Acknowledgement

This work was partially supported by National Natural Science Foundation of China (grant: 61871461, 62171388, 61571374).

### Statements and Declarations

**Competing Interests**: The authors declare that they have no known competing financial interests or personal relationships that could have appeared to influence the work reported in this paper.